\documentclass[a4paper,twocolumn,citeautoscript,reprint]{revtex4-1}

\usepackage{amsmath,bm}
\usepackage[pdftex]{graphicx,color}
\usepackage{epstopdf}
\usepackage[english]{babel}
\usepackage{hyperref}

\usepackage{tikz}
\newcommand*\circled[1]{\tikz[baseline=(char.base)]{
            \node[shape=circle,draw,inner sep=1pt, fill = yellow] (char) {\textcolor{blue}{\small{#1}}};}}

\begin{document}

\title{Super cavity solitons and the coexistence of multiple nonlinear states in a tristable passive Kerr resonator}

\author{Miles Anderson$^{1}$}
\altaffiliation{Current address: \'Ecole Polytechnique F\'ed\'erale de Lausanne (EPFL), CH-1015 Lausanne, Switzerland}
\author{Yadong Wang$^1$}
\author{Fran\c{c}ois Leo$^{1}$}
\altaffiliation{Current address: Universit\'e libre de Bruxelles, 50 Avenue F. D. Roosevelt, CP 194/5, B-1050 Bruxelles, Belgium}
\author{St\'ephane Coen$^1$}
\author{Miro Erkintalo$^1$}
\email{m.erkintalo@auckland.ac.nz}
\author{Stuart G. Murdoch$^1$}
\email{s.murdoch@auckland.ac.nz}

\affiliation{$^1$The Dodd-Walls Centre for Photonic and Quantum Technologies, Department of Physics, The University of Auckland, Auckland 1142, New Zealand}

\begin{abstract}
  \noindent Passive Kerr cavities driven by coherent laser fields display a rich landscape of nonlinear physics, including bistability, pattern formation, and localised dissipative structures (solitons). Their conceptual simplicity has for several decades offered an unprecedented window into nonlinear cavity dynamics, providing insights into numerous systems and applications ranging from all-optical memory devices to microresonator frequency combs. Yet despite the decades of study, a recent theoretical study has surprisingly alluded to an entirely new and unexplored paradigm in the regime where nonlinearly tilted cavity resonances overlap with one another [T. Hansson and S. Wabnitz, J. Opt. Soc. Am. B \textbf{32}, 1259 (2015)]. We have used synchronously driven fiber ring resonators to experimentally access this regime, and observed the rise of new nonlinear dissipative states. Specifically, we have observed, for the first time to the best of our knowledge, the stable coexistence of dissipative (cavity) solitons and extended modulation instability (Turing) patterns, and performed real time measurements that unveil the dynamics of the ensuing nonlinear structures. When operating in the regime of continuous wave \emph{tristability}, we have further observed the coexistence of two distinct cavity soliton states, one of which can be identified as a ``super'' cavity soliton as predicted by Hansson and Wabnitz. Our experimental findings are in excellent agreement with theoretical analyses and numerical simulations of the infinite-dimensional Ikeda map that governs the cavity dynamics. The results from our work reveal that experimental systems can support complex combinations of distinct nonlinear states, and they could have practical implications to future microresonator-based frequency comb sources.
  \end{abstract}
\maketitle
\section{Introduction}
Beginning with theoretical studies of bistability~\cite{szoke_bistable_1969}, the behaviour and dynamics of externally driven nonlinear optical cavities have been extensively investigated for almost 50 years. Besides many application prospects -- ranging from all-optical information storage~\cite{mcdonald_spatial_1990, wabnitz_suppression_1993, barland_cavity_2002, jang_all-optical_2016} to photonic computing~\cite{smith_optical_1984, vinckier_high-performance_2015} --  the continuous interest in such systems stems from the diversity of universal nonlinear physics they support~\cite{lugiato_nonlinear_2015}. Pattern formation and self-organisation~\cite{laughlin_solitary_1983, lugiato_spatial_1987, oppo_formation_1994, arecchi_pattern_1999, rosanov_spatial_2002}, dissipative solitons~\cite{firth_theory_2001,ackemann_chapter_2009, tlidi_introduction:_2007, firth_optical_1996, etrich_solitary_1997, marconi_vectorial_2015}, chaos~\cite{ikeda_optical_1980, carmichael_chaos_1983, mclaughlin_new_1985, garcia-ojalvo_spatiotemporal_2001}, vortices~\cite{coullet_optical_1989, scheuer_optical_1999, gibson_optical_2016}, and topological phase solitons~\cite{garbin_topological_2015} all represent examples of the richness of nonlinear cavity physics.

The simplest nonlinear optical cavity, which nevertheless captures much of the principal dynamics, is arguably that of the passive Kerr cavity. Described as the ``hydrogen atom of nonlinear cavities''~\cite{firth_cavity_2002}, the Kerr cavity model has for several decades offered an unparalleled window into complex cavity dynamics. It has played a particularly ``decisive role in promoting the field of optical pattern formation''~\cite{lugiato_nonlinear_2015}, and in elucidating the intimately related emergence of localised dissipative structures commonly referred to (in optics) as cavity solitons (CSs)~\cite{firth_theory_2001}. Such CSs correspond to localised wave packets that sit on top of a non-zero homogeneous background, and they have been subject to significant research efforts due to their application prospects as bits in all-optical buffers and processing units (for comprehensive reviews, see~\cite{firth_theory_2001, ackemann_chapter_2009, coen_temporal_2016}). Studies focussed initially on spatial CSs~\cite{mcdonald_spatial_1990}, which can manifest themselves as persisting spots in diffractive nonlinear systems, such as semiconductor microcavities~\cite{barland_cavity_2002}. More recently, however, the spotlight has shifted to \emph{dispersive} systems and \emph{temporal} CSs~\cite{wabnitz_suppression_1993, coen_temporal_2016}: pulses of light circulating in optical ring resonators. Whilst first observed~\cite{leo_temporal_2010} and  studied~\cite{leo_dynamics_2013, jang_ultraweak_2013, jang_temporal_2015} in macroscopic resonators constructed from single-mode optical fibers, the interest in temporal CSs has surged over the last couple of years with the identification of their key role in the generation of stable frequency combs in optical microresonators~\cite{herr_temporal_2014, coen_modeling_2013, chembo_spatiotemporal_2013, yi_soliton_2015, brasch_photonic_2016, joshi_thermally_2016, webb_experimental_2016}. Such frequency combs have several potential applications in, e.g., telecommunications and spectroscopy~\cite{kippenberg_microresonator-based_2011, pfeifle_coherent_2014, pfeifle_optimally_2015, suh_microresonator_2016, dutt_-chip_2016}, fuelling continuous efforts to better understand the dynamics that underpin CSs in dispersive Kerr resonators~\cite{milian_soliton_2014, jang_observation_2014, milian_solitons_2015, luo_spontaneous_2015, guo_universal_2016, karpov_raman_2016, yang_stokes_2016, cole_soliton_2016, anderson_observations_2016, bao_observation_2016, yu_breather_2016, lucas_breathing_2016}.

\begin{figure*}[htb]
 \centering
 \includegraphics[width = \textwidth]{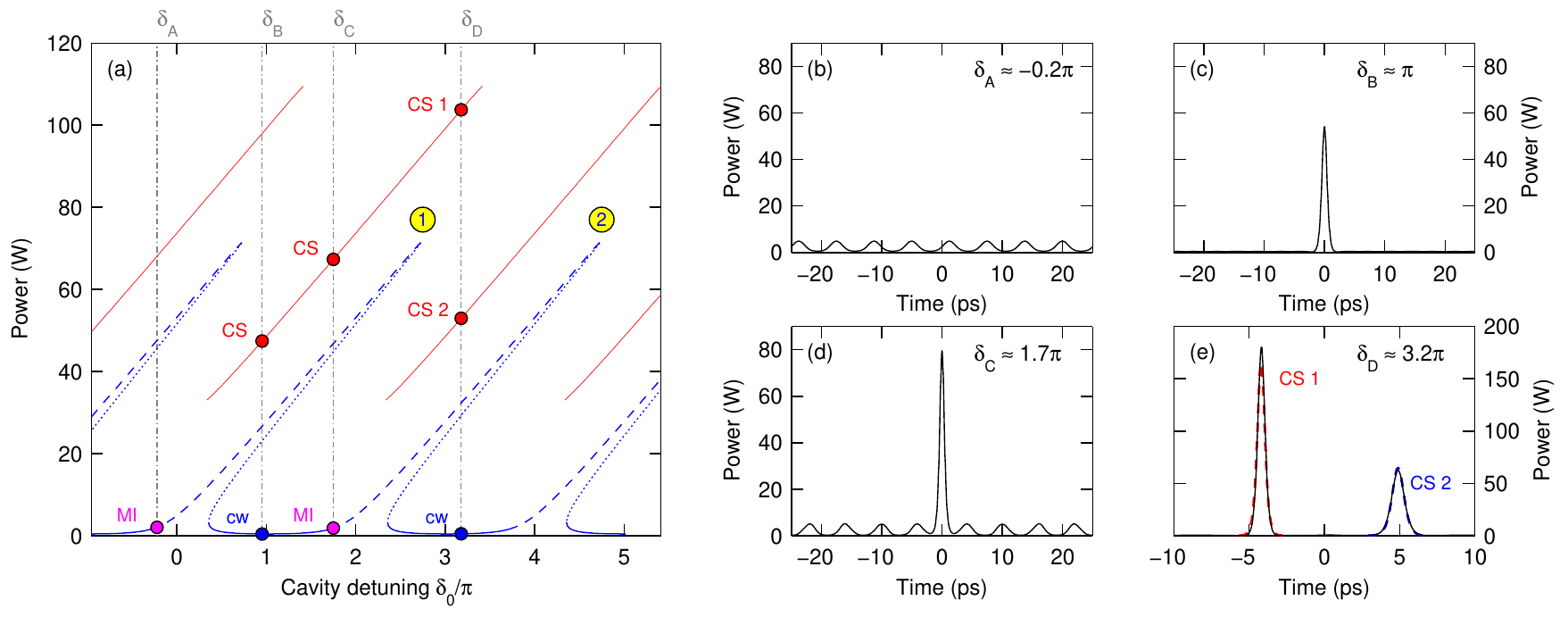}
 \caption{Cavity resonances and examples of nonlinear structures. (a) Blue curves at the bottom show power levels of cw steady-state solutions, while red curves at the top show CS branches predicted individually for the different resonances by the mean-field LLE (see Appendix~\ref{mf}). For clarity, the CS branches are plotted as $f(\delta_0) = 10~\mathrm{W}+P_\mathrm{p}(\delta_0)/2$, where $P_\mathrm{p}(\delta_0)$ is the soliton peak power. (b-e) Examples of nonlinear structures obtained from numerical simulations of the Ikeda map at different detunings (highlighted in (a) as dash-dotted vertical lines): (b) MI pattern, $\delta_0 = -0.7~\mathrm{rad}$; (c) CS on a cw background, $\delta_0 = 3~\mathrm{rad}$; (d) CS coexisting with an MI pattern, $\delta_0 = 5.5~\mathrm{rad}$; (e) coexistence of two CSs associated with adjacent resonances, $\delta_0 = 10~\mathrm{rad}$. In (e), the dashed red and blue curves correspond to CS profiles predicted individually for the different resonances by the LLE. The parameters used in all of the calculations are: $\theta = 0.1$; $P_\mathrm{in} = 15~\mathrm{W}$; $\rho = 0.73$; $\beta_2 = -22~\mathrm{ps^2/km}$; $\gamma = 1.2~\mathrm{W^{-1}km^{-1}}$; $L = 100~\mathrm{m}$. The mean-field results use $\alpha = 0.145$, corresponding approximately to half the total cavity losses per roundtrip (see Appendix~\ref{mf}). Note the different axes in (b)--(d) and (e), highlighting the much larger power and shorter duration of the ``super'' CS.}
 \label{cws}
\end{figure*}

CSs are typically explained to arise under conditions of coexistence between a periodic pattern and a stable homogeneous state~\cite{coen_temporal_2016, lugiato_introduction_2003}; they correspond to localised excitations that connect one cycle of the pattern with the homogeneous state~\cite{parra-rivas_dynamics_2014}. In a pure Kerr cavity, suitable conditions can be readily met in the region of continuous wave (cw) bistability~\cite{coen_universal_2013}, which arises from the tilt of the Lorentzian cavity resonances induced by the Kerr-nonlinear phase shift. But of course, the fact that cavity resonances repeat periodically elicits the question: what if the cavity driving is so strong that the nonlinear phase shift exceeds $2\pi$, i.e., such that adjacent spectral resonances actually overlap [as in Fig.~\ref{cws}(a)]? Despite the decades of study into passive Kerr resonators and CSs, this question has remained virtually unstudied. It is only recently that Hansson and Wabnitz theoretically considered some of the implications of such  strong cavity driving~\cite{hansson_frequency_2015}, motivated by the large phase shifts (of the order $\pi$) already demonstrated in microresonator frequency comb experiments~\cite{delhaye_octave_2011}. Significantly, the authors predicted that, under conditions of cw \emph{tristability}, two different CS states may simultaneously coexist. Because one of the CS states was predicted to possess a significantly shorter duration than the other -- and thus be more attractive for broadband frequency comb generation -- the authors coined the term ``super'' CS for its description~\cite{hansson_frequency_2015}.

In this Article, we report on the first combined experimental and theoretical study of passive Kerr cavity dynamics in the strong-driving regime, as characterised by nonlinear phase shifts in the vicinity of $2\pi$ and beyond. Our experiments are performed using synchronously-driven optical fiber ring resonators, and we observe new nonlinear behaviours emblematic of the strong-driving regime: coexistence of distinct dissipative structures associated with adjacent, nonlinearly-overlapping cavity resonances. In particular, we predict and observe  -- both for the first time to our knowledge -- localised CSs sitting stably atop extended patterned states. Furthermore, we report the first experimental observations of coexisting (super) CSs associated with different characteristics, as predicted earlier by Hansson and Wabnitz~\cite{hansson_frequency_2015}. All our experimental results are in excellent agreement with numerical simulations of the infinite-dimensional Ikeda-map that governs the cavity dynamics, and they conform to a simple physical interpretation of coexistence of nonlinear states associated with individual resonances. Notably, although the full ``mixed'' states we observe are beyond the standard mean-field analysis of passive Kerr cavities~\cite{lugiato_spatial_1987,haelterman_dissipative_1992}, we find remarkably that their constituent nonlinear structures are not.

\section{Theory}
The experiments that will follow are based on an optical fiber ring resonator that is coherently driven by quasi-cw laser light. Several prior studies, performed in the regime of comparatively small nonlinear phase shifts, have demonstrated such systems to be ideal testbeds for the experimental exploration of passive Kerr cavity phenomena~\cite{leo_temporal_2010, leo_dynamics_2013, jang_temporal_2015, jang_controlled_2016, copie_competing_2016, anderson_observations_2016}. We begin by briefly recalling the model equations that govern the system behaviour as well as the steady-state solutions they are known to support. We then argue and demonstrate by means of numerical simulations how new combinations of nonlinear states can emerge when adjacent resonances overlap.

\subsection{Model equations}
The evolution of the slowly-varying electric field envelope in a coherently driven fiber ring resonator is governed by a (generalized) Ikeda map~\cite{ikeda_multiple-valued_1979, coen_modeling_2013, hansson_frequency_2015, haelterman_dissipative_1992, coen_modulational_1997}. In the beginning of each roundtrip, a cw driving field $E_\mathrm{in}$ with power $P_\mathrm{in} = |E_\mathrm{in}|^2$ is coherently superimposed on the lightwave circulating in the resonator, such that the time-domain electric field envelope $E_{m+1}(z,\tau)$ at the beginning of the $(m+1)$th cavity transit obeys the following boundary condition:
\begin{equation}
  \label{boundary}
  E_{m+1}(z=0,\tau) = \sqrt{\theta}\,E_\mathrm{in}+\sqrt{\rho}\,E_{m}(z=L,\tau)e^{-i\delta_0}.
\end{equation}
Here $z$ is the longitudinal coordinate along the optical fiber forming the resonator, $\tau$ is time defined in a reference frame moving at the group velocity of light in the fiber, $\theta$ is the power transmission coefficient of the coupler (located at $z = 0$) used to inject the driving field $E_\mathrm{in}$ into the cavity, $L$ is the roundtrip length of the resonator, and $\delta_0$ is the phase detuning between the driving field and a cavity resonance. For simplicity, we lump all dissipation (arising e.g. from the input coupler, fiber absorption, or component loss along the fiber loop) in the boundary condition, with $1-\rho$ describing the total power lost per roundtrip; the parameter $\rho$ is fully determined by the cavity finesse $\mathcal{F}$, with $\rho\approx1-2\pi/\mathcal{F}$ (valid in the limit $\mathcal{F}\gg 1$). With this approximation, the field envelope $E_{m}(z=L,\tau)$ at the end of the $m$th cavity transit can be obtained by numerically integrating a generalized nonlinear Schr\"odinger equation (NLSE)~\cite{dudley_supercontinuum_2006},
\begin{equation}
\label{GNLSE}
\frac{\partial E_{m}(z,\tau)}{\partial z} = -i\frac{\beta_2}{2}\frac{\partial^2 E_m}{\partial \tau^2 } +  i\gamma \left[R(\tau)\ast|E_m|^2\right]E_m,
\end{equation}
where $\beta_2$ is the group-velocity dispersion coefficient, $\gamma$ is the nonlinearity coefficient, and $R(\tau) = (1-f_\mathrm{R})\delta(\tau) + f_\mathrm{R}h_\mathrm{R}(\tau)$ is the nonlinear response function that includes both the instantaneous Kerr nonlinearity [$\delta(\tau)$ is the Dirac delta function] and stimulated Raman scattering (SRS), with $f_\mathrm{R}$ the Raman fraction of the nonlinearity (for silica glass, $f_\mathrm{R} = 0.18$) and $h_\mathrm{R}(\tau)$ the Raman response function~\cite{stolen_raman_1989}.

As we shall see below, SRS is key to fully explain our experimental observations; however, its role is to merely perturb the nonlinear states supported by pure Kerr cavity dynamics (see Appendix~\ref{SRS}). Accordingly, we begin our discussion by neglecting SRS and set $f_\mathrm{R} = 0$. In this limit, Eqs.~\eqref{boundary} and~\eqref{GNLSE} describe a dispersive cavity with a purely instantaneous Kerr nonlinearity. Because of the equivalence between paraxial-beam diffraction and dispersive pulse spreading~\cite{salem_application_2013}, the system is analogous to a spatially diffractive Kerr cavity, and can thus be considered a generic representation of a one-dimensional Kerr cavity. We remark in this context that, under specific conditions, Eqs.~\eqref{boundary} and~\eqref{GNLSE} can be averaged (see Appendix~\ref{mf}) into a single mean-field equation~\cite{haelterman_dissipative_1992} that is fully analogous to the celebrated Lugiato-Lefever equation (LLE) of spatially diffractive cavities~\cite{lugiato_spatial_1987}. Although our experimental conditions are beyond such a mean-field approximation, the LLE nevertheless provides important insights (as discussed below).

\subsection{Coexistence of multiple nonlinear states}
We are interested in the regime of anomalous dispersion ($\beta_2 < 0$), where the NLSE~\eqref{GNLSE} is self-focussing (for silica fibers, $\gamma > 0$). In this regime, passive Kerr cavities are well-known to support three families of nonlinear states, each of which correspond to a distinct steady-state solution of the Ikeda map~\cite{coen_universal_2013}. These are the {(i) homogeneous} cw states, (ii) periodic (Turing) patterns, and (iii) localised CSs. The different states are closely interrelated: patterned states arise from the modulation instability (MI) of a cw state, while CSs can be understood as combinations of patterned and cw states, corresponding to singular cycles of the pattern sitting atop a cw background~\cite{coen_temporal_2016}. To illustrate how new combinations can emerge when the driving is so strong that adjacent resonances overlap, we plot in Fig.~\ref{cws}(a) the cw steady-state solutions of the Ikeda map (blue curves; see also Appendix~\ref{cwsols}) for parameters similar to the experiments that will follow (listed in the caption of Fig.~\ref{cws}). Here, dotted lines correspond to states that are unconditionally unstable (and will not be considered further), whilst dashed lines highlight states that exhibit MI. Also shown (as red curves) are the CS branches \emph{predicted} for each individual resonance based on the \emph{mean-field} LLE~\cite{leo_temporal_2010, coen_universal_2013}.

The cw solutions in Fig.~\ref{cws}(a) represent periodically repeating cavity resonances that are tilted due to the Kerr nonlinearity (for the parameters used, the Kerr tilt is $\phi_\mathrm{NL}\approx 2.7\pi$). Besides those solutions, inspection of Fig.~\ref{cws}(a) allows us to identify four possible combinations of nonlinear states, and in Figs.~\ref{cws}(b)-(e), we show results from direct numerical simulations of the Ikeda map that illustrate these different behaviours. The simulations all use different cavity detunings [labelled $\delta_\mathrm{A-D}$ in Fig.~\ref{cws}(a)] and initial conditions (MI states grow from random noise whilst CS are excited by assuming an initial condition corresponding to a suitable sech-profile on a cw background). First, referring to resonance \circled{1}, as labelled in Fig.~\ref{cws}(a), MI analysis of the Ikeda map~\cite{coen_modulational_1997} reveals that the upper cw state exhibits MI for cavity detunings {$\delta_1 = \delta_0 \gtrsim \delta_\mathrm{MI} = -0.30\pi$}. Indeed, at the cavity detuning $\delta_\mathrm{A} \approx -0.2\pi$, periodic patterns emerge from an initially noisy background [Fig.~\ref{cws}(b)]. When the detuning increases beyond the up-switching point, which marks the lower boundary of cw bistability, CSs can be expected~\cite{leo_temporal_2010,coen_universal_2013}. For moderate detunings ($\delta_0 < 2\pi$), they correspond to the standard CSs that sit on top of a cw background [Fig.~\ref{cws}(c), $\delta_\mathrm{B} \approx \pi$], with the background coinciding precisely with the stable lower cw state of the resonance (being the only stable cw state available). However, in the strong driving regime, where adjacent resonances overlap, the lower state of the first resonance eventually morphs into the upper state of its neighbouring resonance [labelled \circled{2} in Fig.~\ref{cws}(a)], and can therefore be expected to exhibit MI. Remarkably, as evidenced by the Ikeda map simulation for $\delta_\mathrm{C} \approx 1.7\pi$, CSs continue to exist in this region [Fig.~\ref{cws}(d)]. They now sit, however, on top of a background that is not cw, but rather consists of a periodic MI pattern associated with the upper branch of the second resonance (being the only background state available). Finally, if the driving is sufficiently strong, such that the CS branch from the first resonance extends into the region of cw bistability of the second resonance, a situation may arise where the CS solutions associated with two adjacent resonances can coexist. In this regime, the intracavity field is composed of two different CS states with distinct characteristics (duration, peak power), both of which sit on top of the lower state cw solution of the second resonance [Fig.~\ref{cws}(e), $\delta_\mathrm{D} = 3.2\pi$].

The standard mean-field analysis of passive Kerr cavities -- based on the LLE~\cite{lugiato_spatial_1987, haelterman_dissipative_1992} -- is unable to capture the full mixed nonlinear states associated with overlapping resonances [Figs.~\ref{cws}(d) and (e)]. This can be readily understood by recalling that the cw response of the LLE corresponds to a \emph{unique} Lorentzian resonance [see Appendix~\ref{cwsols}], fundamentally limiting the model's reach to states associated with a single resonance (i.e., MI patterns or CSs atop a cw background). In this context, we re-emphasize that the CS branches shown in Fig.~\ref{cws}(a) were obtained individually for each resonance using the LLE, and should therefore be understood as qualitative predictions only. Somewhat surprisingly, however, we find that, although the full mixed states are beyond the LLE, the constituent nonlinear states are not, despite the large absolute detunings. For example, the dashed red and blue curves in Fig.~\ref{cws}(e) show CS profiles predicted by the LLE for two different cavity detunings, as measured from the centres of the respective linear resonances ($\delta_1 = \delta_0 \approx 3.2\pi$ and $\delta_2 = \delta_1-2\pi\approx 1.2\pi$). As can be seen, the individual profiles predicted by the LLE are in excellent agreement with the CSs that make up the full mixed state obtained from the Ikeda map (black curves). In light of this observation, it is straightforward to explain the CSs' different temporal durations: the LLE predicts the width of a CS to scale as $\Delta\tau\propto\delta_0^{-1/2}$~\cite{coen_universal_2013}, and so the soliton associated with the first resonance expectedly possesses a significantly shorter duration (and broader spectrum) than the one associated with the second resonances. The term ``super'' CS was coined by the authors of~\cite{hansson_frequency_2015} to highlight this difference, yet we emphasize that the two solitons can be understood as the same structures at different detunings [as should be apparent from Fig.~\ref{cws}(a)].

\section{Experimental setups}
To experimentally study the existence of the new mixed nonlinear states, a passive Kerr cavity platform capable of generating large nonlinear phase shifts of the order of $2\pi$ is required. As $\phi_\mathrm{NL} \approx \theta P_\mathrm{in} \gamma L\mathcal{F}^2/\pi^2$, this calls for a strongly-driven cavity with a high finesse and a long roundtrip length. We achieve suitable conditions by using macroscopic fiber ring resonators that are synchronously driven by quasi-cw pulses. Although similar systems have previously been used to successfully study Kerr cavity dynamics~\cite{coen_modulational_1997, copie_competing_2016, anderson_observations_2016}, they have not been optimised to allow access to the highly nonlinear regime where adjacent resonances overlap.

Our experiments have been performed using two different fiber ring resonators, both of which consist of a loop of optical fiber closed on itself with a fiber coupler. The first resonator is similar to the one used in~\cite{jang_ultraweak_2013,jang_temporal_2015}. It is built around a 90/10 coupler, is 100 m long, and (for legacy reasons~\cite{jang_ultraweak_2013,jang_temporal_2015}) incorporates an optical isolator to inhibit stimulated Brillouin scattering (SBS), and a wavelength-division multiplexer which does not play a role in the experiments reported here. The cavity has a total measured finesse of 20 ($\rho\approx 0.73$), which corresponds to 27\% losses per roundtrip. The parameters of this cavity are ideal for the study of coexisting CSs and MI patterns. Indeed, as elaborated below, by locking the driving laser to a fixed detuning from the cavity resonance, we have been able to sustain coexisting CSs and MI patterns for several minutes at a time. The second cavity used in our experiments has been custom-built to allow access to the regime where two distinct CS states can coexist. It uses a 95/5 coupler, has a total length of 300 m, and does not include an isolator nor a wavelength-division multiplexer, thereby yielding a higher finesse of about 48 ($\rho\approx 0.88$). As described below, the high finesse, compounded by the long cavity length, has allowed us to observe the spontaneous excitation of coexisting (super) CSs with different characteristics. Both cavities are composed entirely of standard telecommunication single-mode optical fiber (SMF-28) with nonlinearity and group-velocity dispersion coefficients $\gamma = 1.2~\mathrm{W^{-1} km^{-1}}$ and $\beta_2 = -22~\mathrm{ps^2 km^{-1}}$ (at 1550~nm), respectively.

Each of the cavities is driven by quasi-cw pulses synchronised to their respective roundtrip time. The driving pulses are generated by passing the output of a narrow linewidth, 1550~nm, distributed feedback, cw fiber laser through an intensity modulator, followed by a 2~W Erbium doped fiber amplifier~\cite{anderson_observations_2016}. After spectral filtering to remove the amplified spontaneous emission component of the signal, we obtain flat-top pulses with 4-10 ns duration and a peak power up to 10~W. Thanks to the large peak power of the quasi-cw driving pulses, very large nonlinear phase shifts can be induced. For the 100-m-long cavity, we estimate the maximum phase shift to be around $2.1\pi$; for the 300-m-long cavity, we can reach phase shifts in excess of $4\pi$. At this point, we note that the driving pulses are sufficiently short to mitigate the detrimental effects of SBS, and we have not observed any signatures of SBS in our experiments. In fact, this represents a key feature in our experiments: the high-finesse of our 300-m-long cavity is precisely underpinned by the absence of an optical isolator that is required for SBS suppression under pure cw driving~\cite{leo_temporal_2010}.

To directly monitor the Kerr cavity dynamics, both resonators include a 99/1 tap coupler whose 1\% output port permits direct measurement of the intracavity field. This field is characterised using either an optical spectrum analyzer, frequency-resolved optical gating (FROG), or a real-time measurement system consisting of a 12.5~GHz amplified photodiode and a 40~GS/s real-time oscilloscope. At about 50~ps, the temporal resolution of this real-time measurement system is significantly longer than the picosecond timescales of the CS and MI patterns we wish to observe. Nonetheless, as we shall see below, provided that individual CSs remain separated from one another by more than this resolution, it is possible to clearly observe the system's dynamical evolution and to infer the emergence of new nonlinear mixed states.

\section{Results}
\subsection{Coexistence of CSs and MI patterns}
We first describe results obtained using our 100-m-long cavity, namely the observation of temporal CSs sitting on a periodic MI pattern [as in Fig.~\ref{cws}(d)]. In these experiments, we set the pump power of the nanosecond driving pulses to about 10~W, which induces a nonlinear Kerr tilt of about $2.1\pi$, thus permitting the overlap of two consecutive resonances. To stabilise the phase detuning $\delta_0$, we use a proportional-integral-derivative (PID) servo system that monitors the average power exiting the resonator at the 99/1 tap coupler (locking the average output power to a set level locks the detuning~\cite{jang_ultraweak_2013}).

\begin{figure}[htb]
 \centering
 \includegraphics[width = \columnwidth]{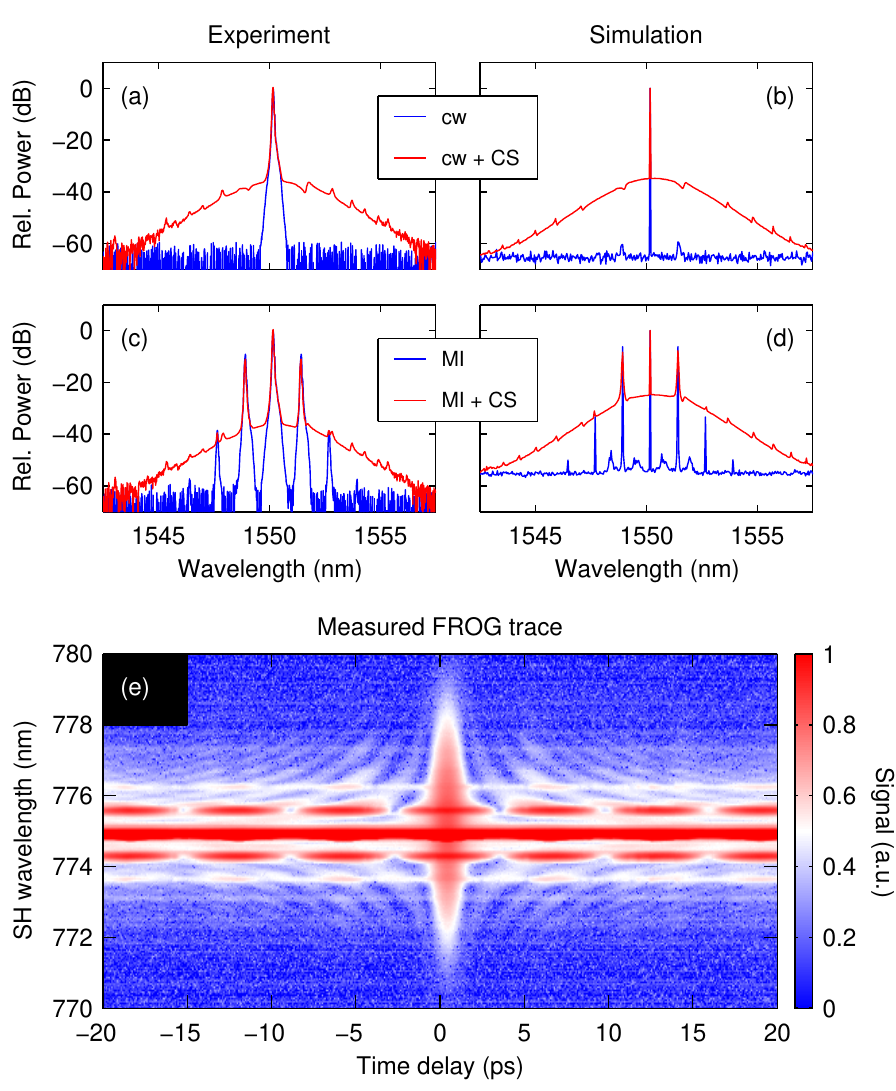}
 \caption{Experimental evidence of coexisting CSs and MI patterns. (a) Experimentally measured spectra at the cavity output before (blue curve) and after (red curve) the excitation of a CS, with the cavity detuning set below the MI threshold of the second resonance ($\delta_1 \approx 1.71\pi; \delta_2 \approx -0.29\pi)$. (b) Corresponding numerical simulations. (c) Same as (a) but with the detuning set above the MI threshold of the second resonance ($\delta_1 \approx 1.79\pi; \delta_2 \approx -0.21\pi)$. (d) Numerical simulations corresponding to (c). (e) Experimentally measured FROG trace corresponding to the spectrum shown in (c).}
 \label{results1}
\end{figure}

MI analysis of the Ikeda map~\cite{coen_modulational_1997} predicts that, for our experimental conditions, an individual resonance will support patterned MI states for detunings larger than a threshold value of $\delta_{2} \approx -0.25\pi$, corresponding to $\delta_{1} = 2\pi + \delta_{2} \approx 1.75\pi$ relative to the preceding resonance. For the sake of discussion, we start by demonstrating the standard configuration of a CS sitting atop a cw background [as in Fig.~\ref{cws}(c)], and lock the pump detuning just \emph{below} the MI threshold, at $\delta_2 \approx -0.29\pi$ ($\delta_1 \approx 1.71\pi$). The blue curve in Fig.~\ref{results1}(a) shows the optical spectrum measured at the output of the 99/1 tap coupler \emph{before} a CS is excited. As can be seen, the spectrum is composed of a single component at the pump wavelength, evidencing a quasi-cw intracavity field. We then excite a temporal CS corresponding to the first resonance (with detuning $\delta_1 = 1.71\pi$) by abruptly perturbing the system. Specifically, by cycling the sign of the proportional component of the PID's output, we rapidly (within 100~ms) sweep the cavity detuning towards the zero of the first resonance and then back to its original value. (As in~\cite{herr_temporal_2014, luo_spontaneous_2015}, CSs are excited as the detuning is scanned back to the original value.) After this perturbation, the spectrum measured at the cavity output [red curve in Fig.~\ref{results1}(a)] clearly shows a broad sech-shaped feature indicative of a temporal CS, superimposed on top of the original quasi-cw field. Figure~\ref{results1}(b) shows corresponding spectra obtained from numerical simulations (parameters as quoted above), and we can see very good agreement with experimental observations. (This simulation, and all the simulations that follow, were obtained from Eqs.~\eqref{boundary} and~\eqref{GNLSE} using experimental parameters quoted previously, and with SRS included with $f_\mathrm{R} = 0.18$.)  In this context, we note that the small dips and peaks in the CS spectrum (visible in both measurements and simulations) correspond to Kelly-like sidebands that arise from the periodicity of the cavity~\cite{jang_observation_2014, kelly_characteristic_1992, wang_real_2016}.

To now demonstrate the new combination of a CS and an MI pattern [as predicted in Fig.~\ref{cws}(d)], we repeat the experiment above but with the cavity detuning locked just \emph{above} the threshold of MI of the second resonance at $\delta_2 \approx -0.21\pi$ ($\delta_1 \approx 1.79\pi$). Similar to Fig.~\ref{results1}(a), the blue curve in Fig.~\ref{results1}(c) shows the measured spectrum in the absence of CSs. In agreement with the prediction that no stable cw state should exist, we see clear spectral signatures of an MI pattern: sidebands equally spaced by 160~GHz. We then excite a CS using the same detuning-sweep approach as above, and the red curve in Fig.~\ref{results1}(c) shows the spectrum measured after the perturbation. Remarkably, we again see the broad sech-shaped feature characteristic of a CS, but now superimposed on top of the original MI pattern. Figure~\ref{results1}(d) shows the spectrum of a numerically simulated state corresponding to a CS atop a MI pattern, and we see very good agreement with experimental observations. In addition to spectral measurements, we have also recorded the FROG trace of this state. This is shown in Fig.~\ref{results1}(e), and consists of two components: a temporally extended, modulated structure with a period of 6.3~ps, and an isolated picosecond pulse centered around the zero delay. The trace is clearly consistent with an intracavity field akin to that shown in Fig.~\ref{cws}(d), i.e.,  a picosecond-scale temporal CS surrounded by a 160~GHz MI pattern. Our experiments show that this state persists as long as the cavity detuning stays locked, which typically corresponds to a timescale of several minutes (equivalent to hundreds of millions of photon lifetimes).

\begin{figure}[htb]
 \centering
 \includegraphics[width = \columnwidth]{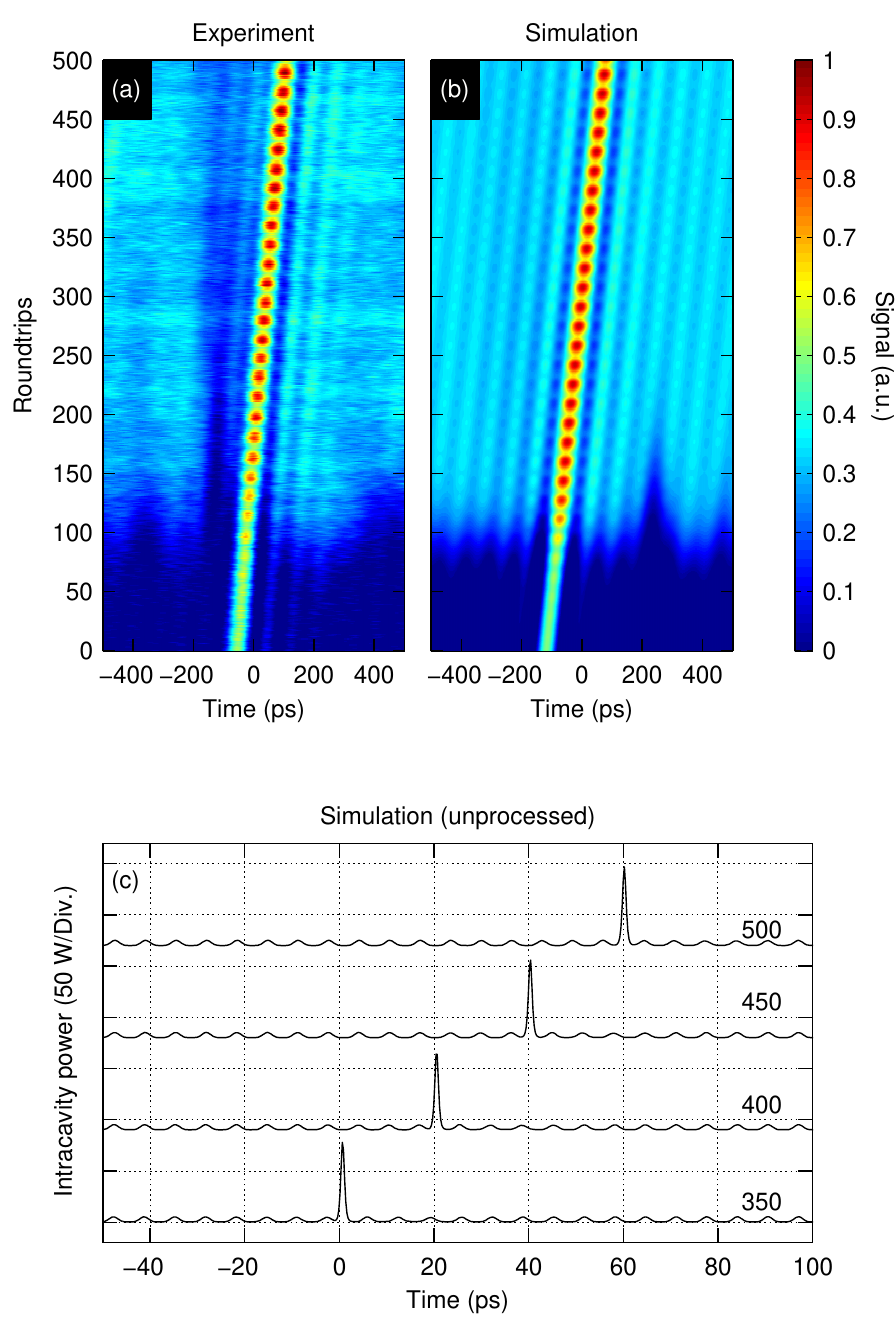}
 \caption{Real time dynamics of a CS on top of a patterned background. (a) Vertically concatenated segments extracted from a single oscilloscope trace measured at the cavity output after an off-set filter, showing the roundtrip-by-roundtrip evolution of the intracavity field after the detuning has settled to a constant value. (b) Corresponding results from numerical simulations, processed to take into account the experimental detection method. The oscillatory features arise from the CS drifting across the MI pattern. (c) Temporal profiles at four different roundtrips (indicated on the right) extracted from the simulation before taking into account the experimental detection method. The profiles are vertically offset for clarity. For a full animation of the simulation, see supplementary material~\cite{sup1}.}
 \label{results2}
\end{figure}

To further confirm that the results in Fig.~\ref{results1}(c) and (e) correspond to a genuine mixed state, where the CS sits directly atop an MI pattern, we have measured the time-resolved roundtrip-by-roundtrip dynamics of the intracavity field. In order to clearly distinguish between a cw background and an MI pattern (whose 160~GHz repetition rate is beyond the bandwidth of our detectors), we record the output field after a 1.6~nm (full width at half maximum) bandpass filter that is offset by about 1.5~nm from the pump wavelength. This removes the cw component at the pump wavelength, thus ensuring that all measured signals are purely borne of the MI and CS fields.

The density map in Fig.~\ref{results2}(a) shows a sequence of experimentally recorded oscilloscope traces, concatenated on top of each other so as to illustrate the roundtrip-by-roundtrip evolution of the intracavity field. The measurement has been taken immediately after the CS excitation process, when the detuning sweep has settled to the constant value of $\delta_1 = 1.79\pi$, and it captures the nascent MI pattern (extended light blue trace) emerging from the cw background (zero signal, dark blue) that initially surrounds a single CS. Surprisingly, as soon as the MI pattern emerges, the CS signal begins to noticeably oscillate. These oscillations (alongside with the other surrounding dynamics), are fully captured by numerical simulations of our experiment, as shown in Fig.~\ref{results2}(b). To better corroborate our experimental findings, the simulation results shown here have been post-processed to mimic our detection methods. Closer analysis of the unprocessed data [c.f. Fig.~\ref{results2}(c)] reveals that the oscillations originate from the group-velocity mismatch between the CS and the MI pattern, dominantly caused by the soliton's Raman self-frequency shift~\cite{milian_solitons_2015, yi_soliton_2015, karpov_raman_2016, anderson_measurement_2016}. Specifically, SRS shifts the spectral centre of the soliton towards longer wavelengths, which gives rise to a change in group velocity (see Appendix~\ref{SRS}); the oscillations in power measured through the offset filter arise as the CS drifts across the periodic MI pattern (see Fig.~\ref{results2}(c) and supplementary animation~\cite{sup1}). Given that the temporal period of the measured pattern is about 6.3~ps, and that the oscillations have a constant period of 16.1 roundtrips, we experimentally estimate a CS drift rate of $V = 0.39~\mathrm{ps/roundtrip}$. This, in turn, corresponds to a Raman self-frequency shift of $\Delta f = V/(2\pi\beta_2) \sim -27~\mathrm{GHz}$, in close agreement with the subtle $25~\mathrm{GHz}$ redshift inferred from the experimentally measured spectrum shown in Fig.~\ref{results1}(c). Both of these values are also in excellent agreement with those extracted from our simulations: $V_\mathrm{sim} \sim 0.40~\mathrm{ps/roundtrip}$ and $\Delta f_\mathrm{sim} \sim -28~\mathrm{GHz}$.

The experimental results shown in Fig.~\ref{results1} and Fig.~\ref{results2} very clearly confirm that the states we have observed correspond to a genuine mixed state where a CS is surrounded by an extended MI pattern, and this interpretation is fully corroborated by our numerical simulations. In addition to representing the first experimental observation of such a state, it is also worth emphasizing that, to the best of our knowledge, the coexistence of MI patterns and CSs associated with adjacent resonances has not even been theoretically proposed before.

\subsection{Coexistence of distinct CS states}
The 100-m-long cavity, used in the experiments described above, does not easily permit nonlinear phase shifts sufficient for two adjacent resonances to simultaneously support temporal CSs. To access that regime, we use our 300-m-long, high-finesse cavity. We set the peak power of the flat-top driving pulses to about 2.6~W, which generates a Kerr tilt of $3.7\pi$. For these parameters, the cavity exhibits full cw tristability for detunings $\delta_2 > 0.22\pi$ ($\delta_1 > 2.22\pi$) measured from the second (first) resonance, and is thus expected to display signatures of coexisting CS states.

\begin{figure}[htb]
 \centering
 \includegraphics[width = \columnwidth]{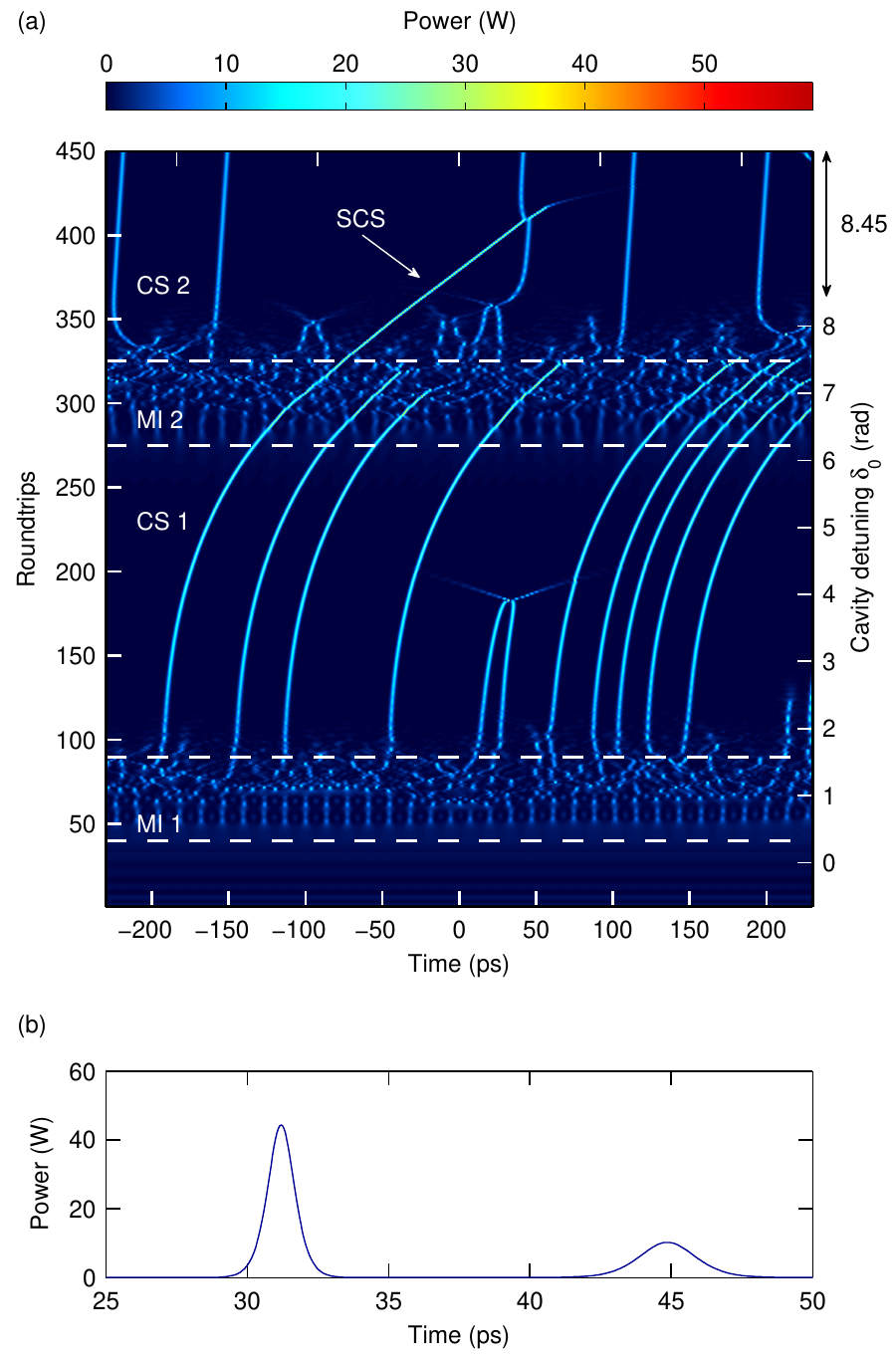}
 \caption{(a) Numerical simulation of the Ikeda map, showing the intracavity dynamics as the cavity detuning is scanned across two cavity resonances and then maintained at $\delta_0 = 8.45~\mathrm{rad}$. Dashed horizontal lines highlight the boundaries of different dynamical regimes, as indicated. (b) Simulated temporal profile at roundtrip 400, before the collision of the two CSs associated with different resonances. SCS: ``super'' CS associated with the first resonance. For a full animation of the simulation, see supplementary material~\cite{sup2}.}
 \label{SCSsims}
\end{figure}

Numerical simulations of the Ikeda map show that, for large detunings, the direct excitation~\cite{leo_temporal_2010, jang_writing_2015} of a CS is extremely difficult, requiring a very carefully shaped initial condition. As it is not feasible to experimentally tailor a perturbation with sufficient precision, we rely on the spontaneous excitation and adiabatic transformation of CSs as the cavity detuning is continuously increased~\cite{herr_temporal_2014, luo_spontaneous_2015}. Specifically, to reach an intracavity state consisting of two distinct CSs atop a cw background [as in Fig.~\ref{cws}(e)], we scan the frequency of the pump laser across two cavity resonances. To facilitate the interpretation of our experimental data, we first describe results obtained from corresponding numerical simulations of Eqs.~\eqref{boundary} and Eqs.~\eqref{GNLSE}. For these simulations, we set the initial detuning to $\delta_0 = -0.65~\mathrm{rad}$, continuously increase it to $\delta_0=8.45~\mathrm{rad}$ over 350 roundtrips, then subsequently maintain it at this level for 100 further roundtrips. Results are shown in Fig.~\ref{SCSsims}, where we plot the simulated evolution of the intracavity intensity over a 500~ps time window as a function of roundtrip number and detuning.

In our simulation, we first see the formation of an MI pattern across the entire cavity [roundtrips 50 -- 85; labelled ``MI 1'' in Fig.~\ref{SCSsims}(a)] as we scan along the first resonance. As expected based on earlier studies~\cite{coen_universal_2013}, the pattern is initially stable, but then transforms into an unstable state consisting of fluctuating structures. Around the 100th roundtrip, a sequence of localised temporal CSs is seen to emerge from the unstable MI state, evolving freely until the detuning reaches the second resonance [roundtrips 85 -- 270; ``CS 1'' in Fig.~\ref{SCSsims}(a)]. The CSs occasionally collide with one another, which leads to merging or annihilation depending on the precise detuning at the roundtrip of collision~\cite{luo_spontaneous_2015, jang_controlled_2016}.  We can also see how the CSs exhibit markedly curved trajectories. This is due to the combined effect of SRS, GVD, and the continuously increasing detuning. Specifically, the CS drift rate is proportional to the Raman-induced redshift ($V \propto \beta_2 \Delta f$), which has been shown to be quadratically proportional to the detuning ($\Delta f \propto \delta_0^2$)\cite{anderson_measurement_2016}. At roundtrip 270, the driving laser passes the MI threshold of the second resonance, and we can indeed see the emergence of a periodic pattern that coexists with the CSs from the first resonance [``MI 2'' in Fig.~\ref{SCSsims}(a)]. As before, the MI pattern is initially stable but develops strong roundtrip-to-roundtrip fluctuations for larger detunings.

Our simulations show that the CSs emerging from the first resonance have a high likelihood of disappearing in the unstable MI region of the second resonance. More detailed analysis reveals that, although CSs can stably coexist with stable MI patterns, they tend to annihilate in the unstable MI regime as they collide with sufficiently large background fluctuations. In the simulation shown in Fig.~\ref{SCSsims}(a), only a single temporal CS (highlighted by a white arrow) survives the MI region, persisting as the detuning increases to the regime of cw tristability. In that region, the CS from the first resonance coexists with those reshaping from the unstable MI pattern associated with the second resonance; the former now corresponds to a ``super'' CS as defined in~\cite{hansson_frequency_2015}. Because that (super) CS is associated with a much larger detuning than those associated with the second resonance, it possesses a significantly shorter duration [as seen in Fig.~\ref{SCSsims}(b)], and as a result, a more pronounced Raman-induced redshift (see also Appendix~\ref{SRS}). This difference in central wavelengths is evident based on the visibly different group-velocities. [We emphasize that the soliton trajectories are linear (instead of curved) in the regime of CS coexistence as the detuning is maintained at a constant value in this region]. Because of their different velocities, the ``super'' CS eventually collides with a soliton associated with the second resonance, and is annihilated due to the perturbation suffered.

In the experimental measurements corresponding to simulations shown in Fig.~\ref{SCSsims}, we monitor the roundtrip-by-roundtrip evolution of the intracavity field as we scan the pump laser across consecutive resonances. Specifically, we record the output of the 99/1 tap coupler using two 12.5~GHz photodetectors that are simultaneously sampled by the 40 GS/s real-time oscilloscope: the first photodiode detects the output field directly, thus recording a signal proportional to the energy of each CS, while the second photodiode measures the output after it has been spectrally filtered by a 1~nm optical bandpass filter offset by 2~nm from the driving laser. Because the spectral width of a CS associated with the first (second) resonance is about 2~nm (1~nm) in the regime where they coexist, our second detection channel responds primarily to the presence of the solitons associated with the first resonance (i.e., the ``super'' CSs). This allows us to unambiguously discriminate experimentally between the two types of nonlinear structures.

\begin{figure}[htb]
 \centering
 \includegraphics[width = \columnwidth]{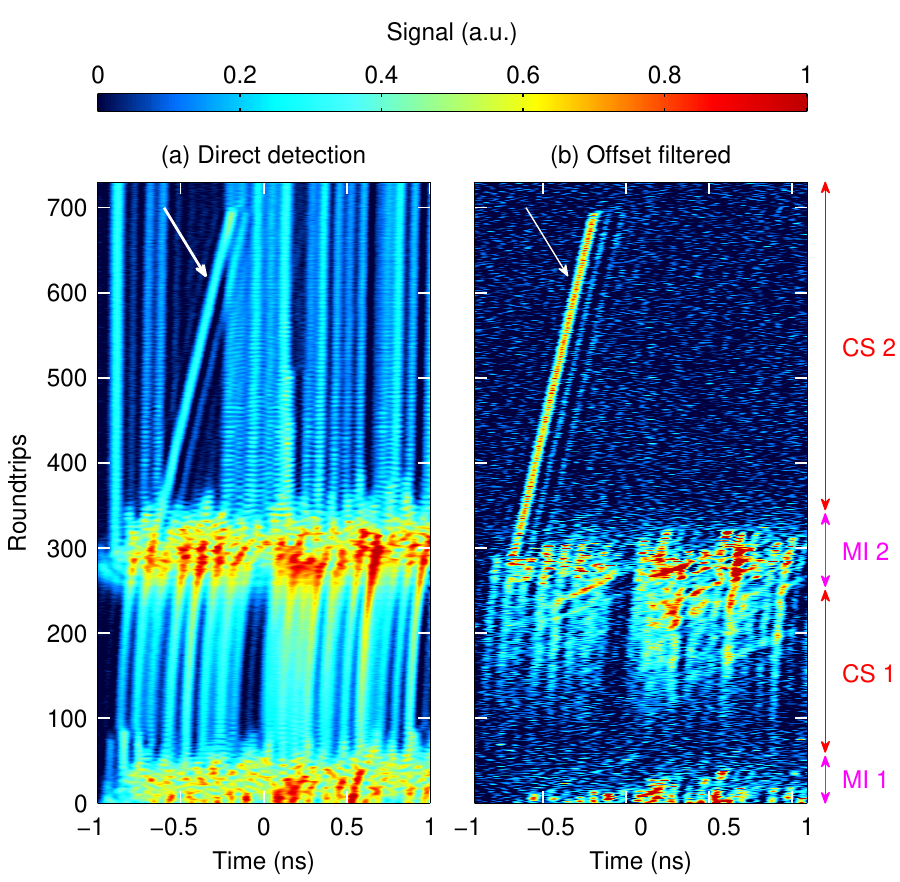}
 \caption{Experimental results, showing evidence of the coexistence of two different CS states. (a, b) Vertically concatenated segments of oscilloscope traces measured as the driving laser frequency is first scanned across two cavity resonances (roundtrips 1--350) and then maintained at a constant value (roundtrips 351 -- 750). Traces in (a) were measured directly at the cavity output whilst those in (b) were obtained after the output had passed through an offset filter. The different slopes and spectral bandwidths of the localized structures emerging around the 350th roundtrip are indicative of two different CS states. The white arrow highlights a ``super'' CS associated with the first resonance.}
 \label{results3}
\end{figure}

In Figs.~\ref{results3}(a) and (b), we show oscilloscope traces recorded by our direct and offset filtered detection channels, respectively. As in our simulations above, the cavity detuning continuously increases until roundtrip 350, after which we allow the system to evolve freely (i.e., without changing or locking the pump cavity detuning~\footnote{Although the detuning is not locked, it can be assumed constant over the 0.5~ms measurement time.}). Despite the limited temporal resolution of our detection system, the experimental oscilloscope traces show all the features predicted in the numerical simulations. Indeed, the intracavity field first corresponds to an unstable MI state, out of which emerges a sequence of CSs that display noticeably curved time-domain trajectories. When the detuning increases beyond the MI threshold of the second resonance (around roundtrip 270), a new (unstable) MI state arises and wipes out most of the CSs associated with the first resonance. However, as highlighted by the white arrow in Fig.~\ref{results3}(a), a single CS survives the MI region unscathed, coexisting with a sequence of newly-formed solitons associated with the second resonance. We can indeed easily distinguish between the two types of CS states based on two clear experimental observations. First, the solitons clearly exhibit different drift velocities: the lone soliton of the first resonance follows its trajectory prior to the MI region, whilst the curvatures of the newly-formed solitons' trajectories are reset. [Note that, as in Fig.~\ref{SCSsims}, the detuning does not change for roundtrips beyond 350, explaining the linear (rather than curved) trajectories.] Second, only the soliton of the first resonance appears in the trace measured in our offset-filtered detection channel [see Fig.~\ref{results3}(b)], evidencing its much larger spectral width compared to the other solitons. Taken together, these observations reveal unequivocally that we have observed the coexistence of different CS states associated with adjacent resonances. Similarly to our numerical simulations, our experiments also show clear evidence (around roundtrip 680) of the ``super'' CS of the first resonance being annihilated after it collides with a soliton associated with the second resonance.

\section{Conclusions}
In summary, we have reported the first combined experimental and theoretical study of Kerr cavity dynamics in the strongly nonlinear regime, where adjacent cavity resonances overlap. We have demonstrated how new combinations of nonlinear states may emerge in this regime, and how they can be understood as mixed states composed of structures associated with individual resonances.  To the best of our knowledge, we have presented the first experimental demonstration of the stable coexistence between a temporal CS and an MI pattern, and reported the first experimental signatures of the coexistence and interactions between two distinct CS states. In this way, our work directly confirms the theoretically predicted existence of ``super'' CSs~\cite{hansson_frequency_2015}. More generally, however, our theoretical and experimental results show that the coexistence of nonlinear structures is not limited to CSs, but encompasses arbitrary combinations of structures associated with adjacent resonances, including periodic patterned states (both stable and unstable). Furthermore, we expect that our general findings will resonate beyond passive Kerr cavities, and that novel mixed states await discovery in other dissipative systems exhibiting tilted homogeneous solutions, such as e.g. quadratically nonlinear optical resonators~\cite{leo_walk-off-induced_2016, hansson_singly_2017}. By identifying and experimentally confirming the general concept of coexisting nonlinear states, our work paves the way for the future study of such dynamics.

Finally, it is interesting to speculate whether the highly nonlinear regime can be relevant to studies of microresonator frequency combs that are currently under intense investigations. Assuming critical coupling and high cavity finesse, the maximum nonlinear phase shift of a cw state can be approximated as $\phi_\mathrm{NL}\approx\gamma P_\mathrm{in} L \mathcal{F}/\pi$. Whilst $\phi_\mathrm{NL}\ll 2\pi$ in most microresonator studies, strong nonlinearities may be required for the generation of ultra-broadband frequency combs. Indeed, a Kerr tilt of about $0.7\pi$ was reached when generating an octave-spanning comb in a $40~\mathrm{\mu m}$-radius silica micro-toroid ($\mathcal{F} \approx 1.2\cdot 10^6$) driven with $P_\mathrm{in}\approx 1~\mathrm{W}$~\cite{delhaye_octave_2011}. To our knowledge, this represents the largest Kerr nonlinear phase shift reported to date in a microresonator experiment, yet we envisage that the continuous push towards broader spectral bandwidths may ultimately motion such systems into the highly nonlinear regime with $\phi_\mathrm{NL} \gtrsim 2\pi$. We therefore conclude that the new nonlinear states observed in our experiment could have implications for the interpretation of future experiments involving broadband microresonator frequency combs.

\begin{acknowledgments}
We acknowledge support from the Marsden Fund and the Rutherford Discovery Fellowships of the Royal Society of New Zealand. We are grateful for useful discussions with Gian-Luca Oppo, Fabio Biancalana, and Matteo Conforti.
\end{acknowledgments}
\appendix
\section{Stimulated Raman scattering}
\label{SRS}
In the context of Kerr cavity dynamics studied in our work, stimulated Raman scattering dominantly manifests itself by shifting the spectral centre of mass of MI patterns and temporal CSs towards longer wavelengths through the so-called intrapulse Raman scattering~\cite{milian_solitons_2015, yi_soliton_2015, karpov_raman_2016}. Denoting the spectral shift as $\Delta f$, one finds that the magnitude of this effect scales as $\Delta f \propto \Delta\tau^{-4}$, where $\Delta \tau$ is the temporal duration of the nonlinear structure~\cite{yi_theory_2016}. Compounded by the fact that the duration of CSs scales as $\delta_0^{-1/2}$, the effect can thus be expected to be particularly important for CSs at high detunings, with $\Delta f\propto \delta_0^{2}$~\cite{anderson_measurement_2016}. Because of dispersion, the frequency shift also gives rise to change in group-velocity, resulting in a temporal drift with respect to the reference frame moving at the velocity of light of the pump wavelength~\cite{anderson_measurement_2016}. Specifically, over a single roundtrip, a spectral shift $\Delta f$ will result in the accrual of an extra group delay given by $V = 2\pi\Delta f\beta_2 L$.

Because, for constant parameters, different nonlinear structures are associated with different temporal durations [see e.g. Fig.~\ref{cws}(e)], they experience different degrees of Raman self-frequency shift. As shown in Figs.~\ref{results2} and~\ref{results3}, the resulting group-velocity variations result in identifiable signatures in real-time measurements, allowing us to reliably discriminate between different nonlinear states.

It is important to emphasize that, in a dissipative cavity system with constant parameters, the nonlinear structures perturbed by SRS typically reach steady-state in the sense that the spectral shift $\Delta f$ and corresponding (per roundtrip) relative group delay $V$ remain constant from roundtrip-to-roundtrip~\cite{milian_solitons_2015}. This should be contrasted with the dynamics of conventional (conservative) solitons in single-pass fiber or waveguide systems, namely continuous spectral red-shift (accompanied by temporal deceleration) along propagation~\cite{dudley_supercontinuum_2006}. In this context, we emphasize that the curved temporal trajectories observed in Figs.~\ref{SCSsims} and~\ref{results3} are simply manifestations of the (adiabatically) increasing detuning ($V\propto\Delta f\propto\delta_0^2$), and should not be confused with deceleration in single-pass systems with constant parameters.

\section{Mean-field approach}
\label{mf}
It is well-known that, if the intracavity field exhibits negligible (linear and nonlinear) evolution over one roundtrip, and if the cavity detuning $\delta_0\ll 1$, the Ikeda map [Eqs.~\eqref{boundary} and Eqs.~\eqref{GNLSE}] can be averaged into a single externally-driven nonlinear Schr\"odinger equation~\cite{haelterman_dissipative_1992, coen_modeling_2013,karpov_raman_2016,milian_solitons_2015}:
\begin{equation}
\begin{split}
t_\textrm{R}\frac{\partial E(t,\tau)}{\partial t}=&\left[-\alpha-i\delta_0-\frac{iL\beta_2}{2}\frac{\partial^2}{\partial\tau^2}\right]E+\sqrt{\theta}E_\textrm{in}\\
&+i\gamma L\left[R(\tau)\ast|E(\tau)|^2\right]E,
\end{split}
\label{lle}
\end{equation}
where $\alpha \approx (1-\rho)/2$. In the regime of anomalous dispersion ($\beta_2 < 0$), focussing Kerr nonlinearity ($\gamma > 0$), and negligible Raman nonlinearity ($f_\mathrm{R} = 0$), Eq.~\eqref{lle} is fully analogous to the Lugiato-Lefever equation of spatially diffractive Kerr cavities~\cite{lugiato_spatial_1987}.

Because the standard derivation of Eq.~\eqref{lle} assumes $\delta_0 \ll 1$~\cite{haelterman_dissipative_1992}, it is not particularly surprising that the equation cannot describe the mixed states involving several adjacent resonances with $\delta_0\approx 2\pi$ [see also discussion on cw solutions in Appendix~\ref{cwsols}]. However, as shown in Fig.~\ref{cws}(e), we generically find that the individual nonlinear states (CSs, patterns) that make up the mixed states are quite well reproduced by Eq.~\eqref{lle}, provided that the cavity detuning $\delta_0$ is quoted relative to the appropriate resonance (and that the cavity has high finesse so as to ensure that the intracavity field evolves only slightly over one roundtrip). This result is somewhat surprising, as it demonstrates the ability of the mean-field Eq.~\eqref{lle} to accurately predict the characteristics of CSs even at very large detunings $\delta_0 \gg 1$. In this context, we emphasize that, although the CS branches shown in Fig.~\ref{cws}(a) were obtained by applying an iterative Newton-Raphson algorithm~\cite{coen_modeling_2013} on Eq.~\eqref{lle}, full simulations of the Ikeda map predict similar ranges of CS existence (with discrepancies arising mostly from the finite finesse).

\section{cw steady-state solutions}
\label{cwsols}
The cw ($\partial E_m(z,\tau)/\partial\tau = 0$) steady-state [$E_{m+1}(z=0) = E_{m}(z=0)$] solutions of the Ikeda map [Eqs.~\eqref{boundary} and Eqs.~\eqref{GNLSE}] satisfy the familiar Airy equation of a nonlinear Fabry-Perot resonator,
\begin{equation}
\label{resonances}
P = \frac{\theta P_\mathrm{in}}{(1-\sqrt{\rho})^2\left[1+F\sin^2(\frac{\delta_0-\gamma L P}{2})\right]},
\end{equation}
where $P = |E_{m}(z=0)|^2$ and $P_\mathrm{in} = |E_\mathrm{in}|^2$ correspond to power levels of the intracavity and the driving fields, respectively, and $F = 4\sqrt{\rho}/(1-\sqrt{\rho})^2$. The solutions of Eq.~\eqref{resonances} describe the periodically repeating, tilted cavity resonances, whose peaks are nonlinearly displaced by $\phi_\mathrm{NL} = \gamma L P_\mathrm{max}$, where $P_\mathrm{max} = \theta P_\mathrm{in}/(1-\sqrt{\rho})^{1/2}\approx\theta P_\mathrm{in}\mathcal{F}^2/\pi^2$ is the peak intracavity power. When the maximum phase displacement $\phi_\mathrm{NL}$ is larger than the resonance width $\Delta\phi = 2\pi/\mathcal{F}$, the cw response becomes multi-valued and exhibits the well-known hysteresis of dispersive optical bistability. In the more extreme situation, where $\phi_\mathrm{NL} > 2\pi$ [as in Fig.~\ref{cws}(a)], adjacent resonances actually overlap. In addition to bistability associated with individual resonances, such a large nonlinear tilt can give rise to regions of cw \emph{tristability}, i.e., regions where the system has three homogeneous equilibrium points that are stable against cw perturbations~\cite{hansson_frequency_2015}.

\begin{figure}[htb]
 \centering
 \includegraphics[width = \columnwidth]{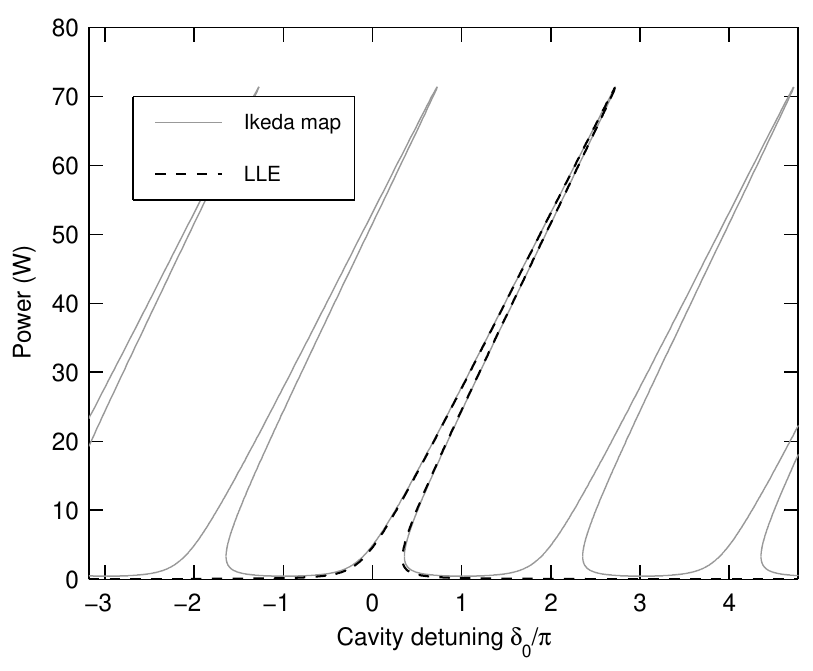}
 \caption{Comparison of cw solutions of the Ikeda map (solid gray curves) and the LLE (dashed black curves). All parameters the same as in Fig.~\ref{cws}.}
 \label{Acws}
\end{figure}

In the mean-field limit, where the Kerr cavity dynamics are described by Eq.~\eqref{lle}, the cw steady-state solutions satisfy the well-known cubic polynomial:
\begin{equation}
\label{mf_resonances}
\theta P_\mathrm{in} = (\gamma L)^2 P^3 - 2\delta_0\gamma L P^2 + (\alpha^2 + \delta_0^2)P.
\end{equation}
In contrast to Eq.~\eqref{resonances}, the solutions of Eq.~\eqref{mf_resonances} describe a single nonlinearly-tilted Lorentzian resonance. To illustrate the difference, in Fig.~\ref{Acws} we compare the cw solutions obtained from the two different models (parameters as in Fig.~\ref{cws}). As can be seen, the cw solutions obtained from Eq.~\eqref{mf_resonances} (black dashed lines) agree very well with a single cycle of the periodically repeating resonances predicted by Eq.~\eqref{resonances} (gray solid curves), but decay to zero as $\delta_0\rightarrow\pm\infty$.  This should make clear why the mean-field Eq.~\eqref{lle} is unable to describe mixed states consisting of structures associated with adjacent resonances.

\end{document}